\documentclass[a4paper,fleqn,usenatbib]{mnras}

\usepackage{newtxtext,newtxmath}

\usepackage[T1]{fontenc}
\usepackage{ae,aecompl}

\usepackage{graphicx}	
\usepackage{amsmath}	
\usepackage{amssymb}	
\usepackage{stfloats}

\title[MESA Models for epsilon Aurigae]{MESA Models of the Evolutionary State of the Interacting Binary epsilon Aurigae}

\author[Gibson and Stencel]{
Justus L. Gibson,$^{1}$
Robert E. Stencel,$^{1}$
\\
$^{1}$Department of Physics and Astronomy, University of Denver, 2112 E Wesley Ave, Denver 80208, USA\\
}

\date{ \bf \today ... Accepted 6 March 2018. Received 12 February 2018; in original form 8 January 2018}

\pubyear{2018}

\begin{document}
\label{firstpage}
\pagerange{\pageref{firstpage}--\pageref{lastpage}}
\maketitle

\begin{abstract}
Using MESA code (Modules for Experiments in Stellar Astrophysics, version 9575), an evaluation was made of the evolutionary state of the epsilon Aurigae binary system (HD 31964, F0Iap + disk).  We sought to satisfy several observational constraints: 1) requiring evolutionary tracks to pass close to the current temperature and luminosity of the primary star; 2) obtaining a period near the observed value of 27.1 years; 3) matching a mass function of 3.0; 4) concurrent Roche lobe overflow and mass transfer; 5) an isotopic ratio $^{12}C$/$^{13}C$ = 5 and, (6) matching the interferometrically determined angular diameter.  A MESA model starting with binary masses of 9.85 + 4.5 M$_{\odot}$, with a 100 day initial period, produces a 1.2 + 10.6 M$_{\odot}$ result having a 547 day period, and a single digit $^{12}C$/$^{13}C$ ratio.  These values were reached near an age of 20 Myr, when the donor star comes close to the observed luminosity and temperature for epsilon Aurigae A, as a post-RGB/pre-AGB star.  Contemporaneously, the accretor then appears as an upper main sequence, early B-type star.  This benchmark model can provide a basis for further exploration of this interacting binary, and other long period binary stars.
\end{abstract}




\section{Introduction}


The goal of this work was to identify possible evolutionary states for the interacting binary, epsilon Aurigae. Progress in observational parameter determination \citep{St14} made it timely to reconsider the evolutionary status for this long period, eclipsing binary star.  We compare new observational constraints with the time-dependent results of MESA calculations (Modules for Experiments in Stellar Astrophysics, \citep{Pa15}).  The last careful consideration of this question was provided by \citet{We85}. Webbink's evaluation strongly suggested that the most viable models for the supergiant in the epsilon Aurigae system were either: (a) a relatively massive post-main-sequence star in shell helium burning mode, or (b) a supergiant star contracting toward a white dwarf state, having been stripped of most of its hydrogen-rich envelope by a combination of tidal mass transfer to the secondary component and mass loss by a stellar wind.  We sought to re-evaluate these conclusions in light of advances in stellar evolution theory, and stellar evolution codes capable of binary star calculations.

The larger context is the relationship of epsilon Aurigae to its immediate family of zeta Aurigae binaries, and the still larger set of relatives known as Algol binaries.  Zeta Aurigae binaries are long period (10$^3$ days) systems involving an evolved supergiant star, plus an upper main sequence companion.  Although long studied, remaining questions about the evolution of zeta Aurigae stars were discussed by \citet{Sc96}, and by \citet{Gr15}.  Algols are characterized by a main sequence star in a binary with a higher luminosity companion of lower mass.  This lead to the  recognition of an Algol paradox wherein the lower mass star evolved faster than the higher mass one, contrary to expectation.  The solution required mass exchange, from the formerly more massive star to the formerly less massive star, evidence for which is found among Algols in general.   Discussions of the angular momentum evolution in Algol systems was provided by \citet{Ch06} and by \citet{Ib06}.  Non-conservative mass loss and the formation of circumstellar rings can dramatically alter evolutionary results, as we show here.

One of the major challenges in tackling this problem involves the issue of non-uniqueness in evolutionary models, complicated by the uncertain distance in this case, which allows viable high and low mass solutions for the epsilon Aurigae system.  To simplify this picture, we adopt a representative {\it minimum} distance of 737 parsecs \citep{Kl15}, implying that the primary star has log(L/L$_{\odot}$) equal to at least 4.35. The distance assumption only affects the values of luminosity for each star and not the observed parameters such as temperature, orbital period, isotopes and mass function.  We revisit implications of this distance assumption in our summary.

There is strong evidence for mass transfer in epsilon Aurigae's past and present, due to the presence of the disk and the existence of a mass transfer stream \citep{Gr13}, \citep{Gi16}.  By including an active mass transfer criterion, we can reject any models that may meet the right temperature and luminosity constraints, but show no current mass transfer, period increase, or do not meet the mass function constraint.  One problem with high mass, q = 1 models, is that the evolutionary products predict very high luminosities, and such would be inconsistent with a hidden secondary star that otherwise would tend to disrupt the accretion disk.  Here, we define q equal to accretor mass divided by donor mass.

Additional system facts related to our investigation include: (1) the mass function, f(M); (2) the evidence for Roche lobe overflow (RLOF) and a mass transfer stream, and (3) the determination of a low  $^{12}C$/$^{13}C$ isotopic ratio (seen during eclipse third contact).  The mass function, f(M), for an eclipsing binary (with $\sin{i}$ = 1, 90 degree inclination) is f(M) = m$_2$ $^3$ / (m$_1$+m$_2$)$^2$.  Recently, f(M) values ranging from 3.12 \citep{Mo62} to 2.51 \citep{St10}, have been reported, based on SB1 radial velocity curve fitting.  We adopt 3.0 +/- 0.2 as a working value for this paper.  The presence of a mass transfer stream (discovered by \citet{Gr13}, and detailed by \citet{Gi16}), as seen during third contact, implies Roche lobe overflow from the primary star.  RLOF indicates a recent mass loss episode from the primary star.  Finally, a low isotopic ratio, $^{12}C$/$^{13}C$ = 5 (compared with solar, 89), was reported by \citet{St15}, based on Gemini GNIRS spectra of transient CO.  Such a small ratio can imply a post-red giant, pre-AGB evolutionary state, or could be related to 'super-AGB' evolution \citep{Sz17}. One additional fact we can use is the measured angular diameter of the F supergiant star, 2.22 $\pm$ 0.09 milliarcsec (MIRC-H LDD, Table 7, \citet{Kl15}; see also \citet{Ba18}), implying a diameter of 176 solar radii, as seen at a distance of 737 pc.  Note that this diameter and the oft-stated effective temperature, 7500K $\pm$ 250K, implies log (L/L$_{\odot}$) = 4.35 $\pm$0.15.

\section{Methods}

Modules for Experiments in Stellar Astrophysics (MESA version 9575) is a one-dimensional  open-source stellar evolution code developed by \citet{Pa10}, with wide-ranging applicability to problems involving single and binary stars, neutron stars, black holes, mass transfer, and giant planets. Each MESA module controls a certain aspect of the physics or numerical analysis and each are thread-safe, meaning that multi-core processors can use certain routines from different modules simultaneously. The code can be run on almost any personal computer as long as a Fortran compiler is installed. For a full description how MESA functions and all of the available utilities it has, refer to the original papers \citep{Pa15} and references therein.

We ran a series of 43 binary star evolutionary models using the MESA code covering a range of initial masses, mass ratios, and periods. Results of these runs are detailed in Table 1.  For all models, we adopted the mass transfer scheme of \citet{Ko90} and used the transfer efficiency described by \citet{Ta06}. All but four of our models had a mass-transfer efficiency of 70\%, with the other four having an efficiency of 40\%. We use a metallicity of Z = 0.01 \citep{Ba18} and an initial surface rotation velocity of 2 km/s, typical of most main sequence stars. Doubling the rotational velocity had negligible effects on results. MESA includes a modified Reimers mass loss prescription during RGB/AGB phases.  Initial masses we tested ranged from 1 to 24 solar masses, and mass ratios (q equals accretor mass divided by donor mass) ranged from 1.0 to 0.2.  The initial periods ranged from 2 days up to 4300 days. The initially longer period models tended not to reach Roche Lobe overflow, thus lacked mass transfer events or period changes.  The initial eccentricities were all set to zero, with the exception of two models, which had an initial eccentricity of 0.2. The changes to resulting tracks were small. All models were run from the main-sequence until the evolutionary time step became too small, or the code was unable to converge to an acceptable model of an evolved star. This often occurred when the donor was on the AGB, but a few models ran long enough for the donor to become a white dwarf. 
\\

\begin{table*}
\centering
\caption{Grid of Models} 
{\scriptsize
\begin{tabular}{ccccccc}
\hline
{Model} & {M$_{D}$(0), M$_{D}$(t*)} & {M$_{A}$(0), M$_{A}$(t*)} & {Initial Period} & {Period change} & {Mass Function} & {Evolutionary State} \\
& M$_{\odot}$ & M$_{\odot}$ & Days & P(t)/P(0) & At t* & At t*  \\
\hline
31 Models rejected \\
\hline
HM24 & 24.0 $\rightarrow$ 10.67 & 23.0 $\rightarrow$ 32.33 & 2800 & 1.83 & 18.27 & AGB \\
HM24b & 24.0 $\rightarrow$ 24.0  & 23.0 $\rightarrow$ 23.0 & 1000 & 1.0 & 5.51 &  HB \\
iben985\_2 & 9.85 $\rightarrow$ 9.849 & 9.80 $\rightarrow$ 9.80 & 1000 & 0.99 & 2.44 & AGB \\
bm985\_d2 & 9.85 $\rightarrow$ 3.43 & 9.80 $\rightarrow$ 11.73 & 500 & 0.32 & 7.02 & AGB  \\
bm985\_beta & 9.85 $\rightarrow$ 2.04   & 9.80 $\rightarrow$ 12.14 & 100 & 8.62 & 18.90 & AGB \\
bm985\_delta & 9.85 $\rightarrow$ 1.05 & 9.80 $\rightarrow$ 12.44 & 100 & 0.42 & 10.58 & AGB \\
iben985\_3 & 9.85 $\rightarrow$ 9.85   & 9.7 $\rightarrow$ 9.7 & 1000 & 1.0 & 2.39 & AGB \\
iben985\_4 & 9.85 $\rightarrow$ 7.55   & 9.7 $\rightarrow$ 11.31 & 600 & 1.04 & 4.06 & AGB \\
iben985\_e2 & 9.85 $\rightarrow$ 9.85   & 7.0 $\rightarrow$ 7.0 & 10000 & 1.0 & 1.21 & AGB \\
iben985\_c & 9.85 $\rightarrow$ 9.85   & 4.5 $\rightarrow$ 4.5 & 1000 & 1.0 & 0.44 & AGB \\
iben985\_f & 9.85 $\rightarrow$ 9.85   & 2.0 $\rightarrow$ 2.0 & 1000 & 1.0 & 0.06 & AGB \\
iben\_b2 & 6.95 $\rightarrow$ 6.09 & 6.90 $\rightarrow$ 7.50 & 1000 & 0.99 & 2.28 & AGB \\
iben\_b3 & 6.95 $\rightarrow$ 1.22   & 6.90 $\rightarrow$ 10.91 & 500 & 10.25 & 8.83 & AGB \\
ibenmod\_a & 6.95 $\rightarrow$ 0.68   & 6.90 $\rightarrow$ 11.29 & 2.0 & 34.29 & 10.03 & AGB \\
ibenmod\_b & 6.95 $\rightarrow$ 6.95   & 6.90 $\rightarrow$ 6.90 & 2.0 & 1.0 & 1.71 & RGB \\
ibenmod\_b2 & 6.95 $\rightarrow$ 1.08   & 6.90 $\rightarrow$ 11.01 & 2.0 & 12.37 & 9.12 & AGB \\
binmod1 & 6.10 $\rightarrow$ 5.72   & 5.90 $\rightarrow$ 6.17 & 1000 & 1.03 & 1.66 & AGB \\
binmod2 & 6.10 $\rightarrow$ 4.33   & 5.90 $\rightarrow$ 7.14 & 1000 & 1.35 & 2.77 & AGB \\
binmod3 & 6.10 $\rightarrow$ 5.17  & 5.90 $\rightarrow$ 6.55 & 1000 & 0.99 & 2.05 & AGB \\
binmod4 & 6.10 $\rightarrow$ 5.11   & 5.90 $\rightarrow$ 6.60 & 1000 & 0.99 & 2.09 & AGB \\
pac\_c & 5.0 $\rightarrow$  5.0  & 2.50 $\rightarrow$ 2.50 & 4300 & 1.0 & 0.28 & AGB \\
pac\_b & 5.0 $\rightarrow$ 5.0   & 2.50 $\rightarrow$ 2.50 & 87 & 1.0 & 0.28 & RGB \\
pac\_b2 & 5.0 $\rightarrow$ 0.81   & 2.50 $\rightarrow$ 5.43 & 100 & 3.63 & 4.11 & AGB \\
pac\_a & 5.0 $\rightarrow$ 5.0   & 2.50 $\rightarrow$ 2.5 & 1.5 & 0.76 & 0.28 & RGB \\
iben985\_d & 4.5 $\rightarrow$ 4.5   & 9.85 $\rightarrow$ 9.85 & 1000 & 1.0 & 4.64 & MS \\
solar\_twins & 1.01 $\rightarrow$ 0.57   & 1.0 $\rightarrow$ 1.44 & 1000 & 1.88 & 0.74 & AGB \\
\hline
11 Models rejected \\ -- lacking mass transfer \\
\hline
binmod2b & 12.1 $\rightarrow$ 12.1   & 12.05 $\rightarrow$ 12.05 & 1000 & 1.0 & 3.00 & RGB \\
binmod2c & 12.1 $\rightarrow$ 12.1   & 12.05 $\rightarrow$ 12.05 & 1000 & 1.0 & 3.00 & RGB \\
binmod2c\_2 & 12.1 $\rightarrow$ 12.1   & 12.05 $\rightarrow$ 12.05 & 1000 & 1.0 & 3.00 & RGB \\
binmod2a & 12.1 $\rightarrow$ 12.1   & 11.90 $\rightarrow$ 18.90 & 1000 & 1.0 & 2.93 & HB \\
binmod2aaa & 12.1 $\rightarrow$ 12.1   & 11.90 $\rightarrow$ 18.90 & 1000 & 1.0 & 2.90 & HB\\
binmod2a\_650 & 12.1 $\rightarrow$ 12.1 & 11.85 $\rightarrow$ 11.85 & 650 & 1.0 & 2.90 & RGB \\
binmod2aa & 12.1 $\rightarrow$ 12.1 & 11.80 $\rightarrow$ 11.80 & 1000 & 1.0 & 2.90 & RGB \\
iben985g & 12.0 $\rightarrow$ 12.0 & 7.0 $\rightarrow$ 7.0 & 1000 & 0.95 & 0.95 & HB \\
ibenh & 12.0 $\rightarrow$ 12.0   & 5.0 $\rightarrow$ 5.0 & 1000 & 1.0 & 0.43 & RGB \\
ibeni & 12.0 $\rightarrow$ 12.0   & 5.0 $\rightarrow$ 5.0 & 1000 & 1.0 & 0.43 & RGB\\
iben985e & 9.85 $\rightarrow$ 9.85 & 7.0 $\rightarrow$ 7.0 & 1000 & 1.0 & 1.21 & AGB \\
\hline
7 Better Models \\
\hline
binmod2a100 & 12.1 $\rightarrow$ 2.8   & 11.85 $\rightarrow$ 18.36 & 100 & 5.81 & 13.82 & AGB \\
binmod2a300  & 12.1 $\rightarrow$ 2.86  & 11.85 $\rightarrow$ 18.32 & 300 & 5.56 & 13.71 & AGB \\
*Iben985b & 9.85 $\rightarrow$ 1.18  & 4.5 $\rightarrow$ 10.57 & 100 &	5.47 & 8.55 & SAGB \\
Iben985b\_delta & 9.85 $\rightarrow$ 1.34 & 4.50 $\rightarrow$ 10.88 & 100 & 7.35 & 8.63 & SAGB \\
Iben985 & 9.85 $\rightarrow$ 2.01   & 9.8 $\rightarrow$ 15.29 & 100 & 7.62 & 11.95 & AGB \\
bm985both & 9.85 $\rightarrow$ 1.11   &	9.8 $\rightarrow$ 13.3 & 100 &	5.86 & 11.33 & AGB \\
Iben985f300 & 9.85 $\rightarrow$ 3.12   & 2 $\rightarrow$ 6.71 & 300 & 0.14 & 3.13 & RGB-HB \\
\hline
\end{tabular}
}
\\
t* = closest approach to the observed L$_{donor}$, T$_{donor}$
\\
*Model Iben985b was motivated by review of \citet{Ib85}
\end{table*}

\subsection{Discussion of Mass Transfer Details used by MESA}


The mass transfer efficiency scheme that we adopted consisted of four parameters referred to as $\alpha$, $\beta$, $\delta$, and $\gamma$, as defined by \citet{So97}. Alpha is the fractional mass loss from the vicinity of the donor star. We adopted  $\alpha$ = 0.1 meaning that 90\% of the mass lost from the donor ends up on the accretor star.  Beta is the fractional mass loss from the vicinity of the accreting star. We adopted $\beta$ equal to 0.1, meaning that 10\% of the mass lost from the donor escapes from the system. Delta is the fractional mass lost from the circumbinary coplanar toroid, with a radius equal to $\gamma$$^2$a, where a is the binary semi-major axis. We adopted $\gamma$ equals 1.3 to ensure the circumbinary torus radius would exceed the binary semi-major axis. In general, as $\alpha$, $\beta$, and $\delta$ increase, less donor matter makes it onto the accretor. This idea is summarized by the efficiency parameter, $\eta$, where $\eta$ = 1 - $\alpha$ - $\beta$ - $\delta$.  The next step in the process was to explore the effects of changes to these parameters.  We ran a few models with the lower efficiency of 40\%  where two models have $\delta$ = 0.5, one had $\beta$ = 0.5, and the last one had $\beta$ = $\delta$ = 0.25.  The effect of these parameter changes and their effect on the period change is discussed next.
%

Variation of parameters was studied, using prescriptions documented by \citet{So97}, as follows (their equations B7--B10):

\begin{equation}
	\frac{P}{P_o}=(\frac{q}{q_o})^{3A_w-3}\ (\frac{1+q}{1+q_o})^{1-3B_w}\ (\frac{1+\epsilon q}{1+\epsilon q_o})^{5+3C_w}
\end{equation}

\begin{equation}
	A_w = A \alpha + \gamma\delta 
\end{equation}

\begin{equation}
	B_w = \frac{A \alpha + \beta }{1 - \epsilon}
\end{equation}

\begin{equation}
	C_w = \frac{\gamma\delta(1 - \epsilon) }{\epsilon } + \frac{A \alpha\epsilon }{1 - \epsilon\ } + \frac{\beta\ }{\epsilon(1 - \epsilon\ )}
\end{equation}

	We are using subscript w to reflect the wind law (Soberman et al., eqn. 29), rather than the mysterious subscript 5 in their appendix B (which might refer to the five parameters alpha, beta, delta, gamma, and A, the angular momentum loss efficiency, Soberman et al. eqn. 14).  With these equations it is easy to see how a given change to alpha, beta, delta and gamma can affect the final period of the system.  In general, a decrease to alpha, or a decrease to delta, will result in the period growing.  An increase to beta or gamma will increase the period although, there is some sensitivity to the initial mass ratio.

To illustrate the effect of parameter variation, we considered our better model, Iben985b, with initial donor mass = 9.85 M$\odot$, initial accretor mass equals 4.5 M$\odot$, initial period equal to 100 days, and we determined the change in period when the various parameters are adjusted. The results are shown in Table~\ref{tab:mt_variation} The biggest change can be seen when gamma is adjusted.  However, having gamma less than 1 is unphysical because it would make the circumbinary disc smaller than the binary separation.

\begin{figure} 
\begin{center}
\includegraphics[width=1.0\linewidth] {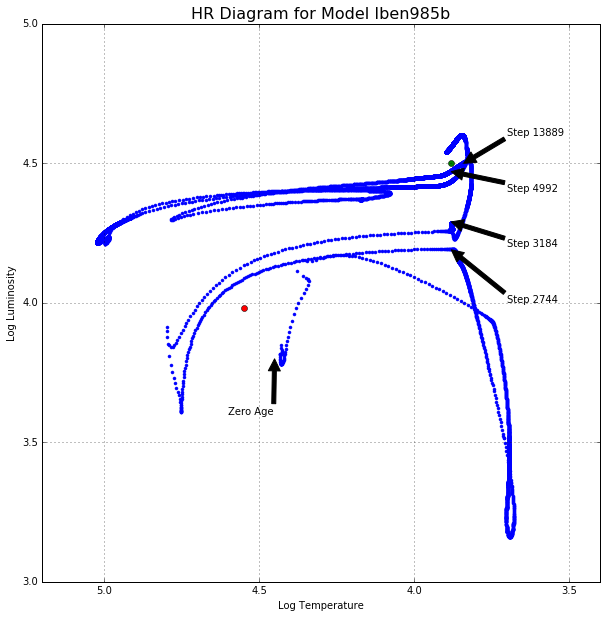}
\caption{The evolutionary paths of our best MESA model, Iben985b, for the epsilon Aurigae binary system. Model step numbers represent time steps in the calculation. The blue lines represent the primary star model steps, with the current observed quantities shown with a green dot near logL = 4.35 and logT = 3.87. Model step 2744 occurs at log age 7.2820 (first ascent red giant branch, 12C/13C = 3.51). Step 3184 occurs at log age 7.3426 (end of horizontal branch, 12C/13C = 3.50). Step 4992 occurs at log age 7.3430 (asymptotic giant branch, carbon/oxygen core, 12C/13C = 24.80). Step 13842 occurs at a log age of 7.3439 (super asymptotic giant branch, oxygen/neon core, 12C/13C = 5.1). The red dot near log L = 4.0, log T = 4.6 indicates the evolved result for the accreting  star (epsilon Aurigae B) during later timesteps in this model. }
\label{fig:HR}
\end{center}
\end{figure}

For each model we ran, we also plotted Hertzsprung-Russell diagrams, as well as plots of mass, radius, period, mass transfer, luminosity, and temperatures, versus both model number and as Kippenhahn diagrams where the x-axis is log(age minus age at end of calculation). In addition to the information these plots provide us, we also looked at the interior structure to determine evolutionary states, and calculated mass ratios and mass functions for an assumed inclination of 90 degrees. For each model we were looking for a period increase, a mass transfer event, a mass function close to the observed value, and a temperature and luminosity for the donor star close to what is observed. For the models that fit these criteria, we found the system age corresponding to the best temperature and luminosity match and then determined the system period, masses, temperatures and luminosities for each star, and evolutionary state for each star at that age.  Our best case was model Iben985b (Fig.~\ref{fig:HR}). 

\begin{table}
\begin{center}
\caption{Variation of Mass Transfer Parameters and the Effect on Period Change}
\label{tab:mt_variation}
{\scriptsize
\begin{tabular}{ccccccc}\hline
{Model} & {A} & {$\alpha$} & {$\beta$} & {$\delta$} & {$\gamma$} & {P(t*)/P(0)} \\
\hline
Iben985b & 1   &	0.1 & 0.1 &	0.1 & 1.3 & 5.76 \\
A/2      & 0.5 &	0.1 & 0.1 &	0.1 & 1.3 & 7.19 \\
A*2      & 2   &	0.1 & 0.1 &	0.1 & 1.3 & 3.69 \\
$\alpha$/2 & 1   &	0.05 & 0.1 &	0.1 & 1.3 & 6.18 \\
$\alpha$*2 & 1   &	0.2 & 0.1 &	0.1 & 1.3 & 5.00 \\
$\beta$/2 & 1   &	0.1 & 0.05 &	0.1 & 1.3 & 5.38 \\
$\beta$*2 & 1   &	0.1 & 0.2 &	0.1 & 1.3 & 6.62 \\
$\delta$/2 & 1   &	0.1 & 0.1 &	0.05 & 1.3 & 9.45 \\
$\delta$*2 & 1   &	0.1 & 0.1 &	0.2 & 1.3 & 2.00 \\
$\gamma$/2 & 1   &	0.1 & 0.1 &	0.1 & 0.65 & 11.11 \\
$\gamma$*2 & 1   &	0.1 & 0.1 &	0.1 & 2.6 & 1.54 \\
\hline
\end{tabular}
}
\end{center}
\end{table}    

\section{Results and Interpretation}

Of our 43 models, 16 had donor star temperatures and luminosities with evolutionary tracks passing within three-tenths dex of the observed primary star luminosity of log L/L$_{\odot}$ = 4.35, and within two-tenths dex of the observed primary star temperature of log T = 3.88.  Of these 16, 10 models can be rejected because they reach the desired donor values at a stage in their evolution where no mass has been transferred and no period change has occurred.  The remaining 7 models, summarized in Table 1, have donor luminosity values within two-tenths of the assumed value and temperature values within one-tenth of the value. Two of the remaining models are 12.1 + 9.85 M$_{\odot}$ systems, with initial periods of 100 and 300 days, that showed increases in their period of 5.8 and 5.6 times, respectively. The other four models have donor masses of 9.85 M$_{\odot}$ with accretor masses of 2, 4.5, and two with 9.8 solar masses. The 9.85 + 2 M$_{\odot}$ case had an initial period of 300 days, while the others had initial periods of 100 days. The difference between the two 9.85+9.8 M$_{\odot}$ with 100 day initial periods is that model Iben985b has a mass transfer efficiency (1 - $\alpha$ - $\beta$ - $\delta$) of 70\% and the iben985b\_delta model has an efficiency of 40\%

\subsection{Models and Alternatives}
\label{sec:maths} 

\begin{figure} 
\begin{center}
\includegraphics[width=1.0\linewidth] {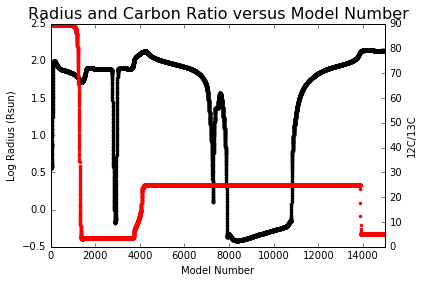}
\caption{Carbon isotopic ratio (red line) and log radius (black line) versus model number for MESA model Iben985b, using the nuclear networks prescribed by \citet{Fa15} - see also \citet{Ka14}.  Model step intervals from $\sim$2000 to 4000 and from $\sim$14000 onward have single digit values of the carbon isotopic ratio, close to the observed value of 5 (see text).    }
\label{fig:RR}
\end{center}
\end{figure}

The results of our best model, Iben985b, suggest that there were two periods in the evolution of the system where the ratio of $^{12}C$/$^{13}C$ have single digit values, as is observed (Fig.~\ref{fig:RR}). These occur between Iben985b step numbers 1000 and 4000, and for step numbers greater than 14000, which correspond to two different evolutionary phases. The advantage of the two best step numbers between 1000 and 4000 (step number 2744 and step number 3184) is in their fit to the mass function and accretor conditions. Step numbers 2744 and 3184 have mass functions of 6.62 and 6.79 respectively, while step number 13889 has a mass function of 8.59. However, the period has only increased by a factor of 1.7, and 1.8 for step numbers 2744 and 3184, but has increased by a factor 5.7 by step number 13889. No models were able to produce a 3 order of magnitude increase to the orbital period necessary to replicate the current 9,890 day period of epsilon Aurigae, but further adjustments to mass transfer parameters could be explored to achieve this. 

\subsection{Predictions}

One thing that we can do with the four best step numbers from Iben985b (Fig.~\ref{fig:HR}) is make predictions about what the epsilon Aurigae system may do in the future, if it is currently represented by any of these step numbers. All of these step numbers have donor radii consistent with the interferometric angular diameter reported by \citet{Kl15}, but different step numbers predict different changes to the radii over different time scales. If the donor star is defined by Iben985b, step number 2744, then MESA predicts that over the next million years, the radius will decrease dramatically from R/R$_{\odot}$ = 10$^{1.86}$ to R$_{\odot}$ $\sim$ 10$^{-0.2}$. Then, over the next million years, the star will begin to expand back to a value of R/R$_{\odot}$ = 10$^{1.9}$.  If instead, the donor star is represented better by step number 3184, then MESA predicts that the star will expand from R/R$_{\odot}$ = 10$^{1.91}$ to 10$^{2.12}$ in roughly 18,000 years. For a donor star represented by step number 4992, MESA predicts the star will shrink drastically from R/R$_{\odot}$ = 10$^2$ to 10$^{0.1}$ over a few hundred years, before expanding back to a value near R/R$_{\odot}$ = 10$^{2.10}$ in 25,000 years. Lastly, if the donor is near step number 13889, then over the next 2000 years, MESA predicts that the star will shrink from R/R$_{\odot}$ = 10$^{2.13}$ to 10$^{2.0}$. Although a long-term decrease in size has been reported by \citet{Sa86}, a study of existing interferometric data by \citet{Kl15} did not verify the earlier claim.  All of this can be summarized by saying that {\it if} the donor star in epsilon Aurigae is currently sitting close to configurations reflected in Iben985b step numbers 2744, 4992, or 13889, then going forward we can expect the star to shrink in radius in the astronomically near-future.  If the star is currently near model 3184 then we can expect the star to expand in the future. 

The state of the accretor for all of the models in Table 2 is more or less the same, with temperatures and luminosities only differing by less than 5\% among individual models. All of these models point to a modern accretor with  L/L$_{\odot}$ $\sim$ 10$^{4.0}$ and a temperature near 30,000K, consistent with an early B-type main-sequence star. Plots of the internal structure of the accretor confirm that it still has a core predominately consisting of hydrogen, due to the substantial mass gain from the donor.

\begin{table} 
\begin{center}
\caption{Model Iben985b vs Constraints} 
{\scriptsize
\resizebox{\columnwidth}{!}{%
\begin{tabular}{cccccccc}
\\
\hline
{Model Step} & {Log Age} & {L$_{Donor}$} & {T$_{Donor}$} & {R$_{Donor}$} & {$^{12}C$/$^{13}C$} & {f(M)} & {P(t)/P(0)} \\
& (t*) & Log L$_{\odot}$ & Log Kelvin & Log R$_{\odot}$ &  &   \\ 
\hline
2744 & 7.2820 & 4.19 & 3.88 & 1.86 & 3.51 & 6.62 & 1.69 \\
3184 & 7.3426 & 4.29 & 3.88 & 1.90 & 3.50 & 6.80 & 1.83 \\
4992 & 7.3430 & 4.47 & 3.88 & 2.00 & 24.80 & 8.59 & 5.65 \\
13889 & 7.3439 & 4.52  & 3.83 & 2.13 & 5.10 & 8.59 & 5.65 \\
Obs. & -- & 4.5$\pm$0.2 & 3.88 & 2.0$\pm$0.2 & 5$\pm$3 & 3.0$\pm$0.3 & -- \\
\hline
\end{tabular}%
}
}
\end{center}
\end{table}   

\section{Conclusions}

We have computed a selection of binary star evolutionary models with MESA code in an effort to examine the relationship of initial versus intermediate states of models with respect to updated observational constraints for the binary star, epsilon Aurigae.  For an assumed distance of 737 pc, we find a model with a primary star starting at 9.85M$_{\odot}$, secondary star starting at 4.5M$_{\odot}$ and initial period of 100 days, to yield a system, after 20 Myr, of an evolved 1.2M$_{\odot}$ star plus a 10.6M$_{\odot}$ upper main sequence star in a 547 day orbital period.  Advantages of this model include: (a) matching presently observed luminosity and effective temperature values; (b) that mass transfer and system mass loss have occurred; (c) that period increase and mass function decrease are in accord with a trend that ultimately may more closely match observed values if certain assumptions in the MESA code were to be modified (e.g. mass loss rates, tidal friction factors \citep{Hu02}, etc.); (d) that a low  $^{12}C$/$^{13}C$ isotopic ratio, as observed, is obtained, and (e) that the measured interferometric diameter can be matched by the luminosity primary. 

Much of the literature associated with this system revolves around the uncertainty in the distance.  We have selected a distance of 737 pc that both encompasses the HIPPARCOS detection of parallax (1.53 $\pm$ 1.29 milli-arcsec) and avoids the impacts of an overly luminous primary and/or secondary star at larger distances.  This is not to dismiss the possibility of a larger system distance, but to establish a way-point on the journey to reconcile how the current interacting binary may have come about.  Given the almost unlimited parameter space open for exploration, our hope was to establish a baseline model that will be open to modification and improvement, based on a new generation of binary star evolutionary codes now available for application. Barring an unlikely capture scenario to create a long period binary, we found support for the role of mass transfer, mass loss, and circumbinary disk formation, to enhance the expansion of an evolving binary system into a zeta Aurigae like system resembling epsilon Aurigae.

\section*{Acknowledgements}

We are grateful for the bequest of William Herschel Womble, in support of Astronomy at the University of Denver, which helped make this study possible. We also thank the reviewer for insightful comments. \\








\appendix

\section{MESA Input Files}

The following blocks of code were run using MESA version 9575. The headings of each block "inlist\_project and "inlist1" refer to the names of the text file used by MESA. We also had another text filed titled "inlist2" which is exactly the same as inlist1 except it is referring to the secondary star. Another file titled "inlist" is also needed to tell MESA how to start. 
\begin{verbatim}
inlist_project
&binary_job
   inlist_names(1) = 'inlist1' 
   inlist_names(2) = 'inlist2'
   evolve_both_stars = .true.
/ ! end of binary_job namelist
&binary_controls
   m1 = 9.85d0  ! donor mass in Msun
   m2 = 4.5d0 ! companion mass in Msun
   initial_period_in_days = 100d
   limit_retention_by_mdot_edd = .false.
   ignore_rlof = .false.
mdot_scheme = 'Kolb'
   mass_transfer_alpha = 0.1d0
   mass_transfer_beta = 0.1d0
   mass_transfer_delta = 0.1d0
   mass_transfer_gamma = 1.3d0
   accretor_overflow_terminate = 15.0d0
   max_tries_to_achieve = 20       
/ ! end of binary_controls namelist

inlist1
&star_job
	mesa_dir = ''
    show_log_description_at_start = .false.
    change_initial_Z = .true.
    new_Z = 0.01
    new_surface_rotation_v = 2 ! (km sec^1)
    set_initial_surface_rotation_v = .true.
    change_net = .true.
    new_net_name = 'sagb_NeNa_MgAl.net'
    adjust_abundances_for_new_isos = .true.
/ ! end of star_job namelist
&controls
    extra_terminal_output_file = 'log1' 
    photo_directory = 'photos1'
    log_directory = 'LOGS1'
    profile_interval = 50
    history_interval = 1
    terminal_interval = 1
    write_header_frequency = 10
    varcontrol_target = 5d-4
/ ! end of controls namelist

\end{verbatim}


\bsp	
\label{lastpage}
\end{document}